\documentstyle[preprint,prl,aps]{revtex}
 
\begin{document}
 \draft
 %\preprint{8-24-93}
 \title{
$SO(5)$ Quantum Nonlinear $\sigma$ Model Theory\\
of the High $T_c$ Superconductivity
 }
 \author{Shou-Cheng Zhang}
 \address{
Department of Physics, Stanford University, Stanford, CA 94305
}
 
\date{\today}

\maketitle
 \begin{abstract}
We  show that the complex phase diagram of high $T_c$ superconductors
can be deduced from a simple symmetry principle,
a $SO(5)$ symmetry  which unifies antiferromagnetism with
$d$ wave superconductivity. We derive the approximate
$SO(5)$ symmetry from the microscopic Hamiltonian and show
furthermore that this symmetry becomes
exact under the renormalization group flow towards a bicritical
point. With the help of this symmetry, we 
construct a $SO(5)$ quantum nonlinear $\sigma$ model to describe
the effective low energy degrees of freedom of the high $T_c$ 
superconductors, and use it to deduce the phase diagram and the
nature of the low lying collective excitations of the system.
We argue that this model naturally
explains the basic phenomenology of the high $T_c$ superconductors
from the insulating to the underdoped and the optimally doped region.

 \end{abstract}
% \pacs{PACS numbers: 7.10.Bg, 71.55.Jv}
 \newpage
 
 %\tighten
 \narrowtext

{\bf In search of a low energy effective theory} 
The high $T_c$ superconductors are among the most complex systems
ever studied in condensed matter physics. In order to reach to 
the core of the physics responsible for this novel phenomenon,
we obviously have to cross many different energy scales, and 
integrate away irrelevant degrees of freedom. Earlier in the
high $T_c$ development, Anderson\cite{anderson}, 
Zhang and Rice\cite{zhangrice} argued rather
successfully that a good starting point for modeling the strong 
correlation effects in the oxides should be the Hubbard, or 
the $t-J$ model close to half filling. Unfortunately, both of
the these models are still too difficult treat. Their exact
solutions in one dimension shed very little light on how to solve
these models in higher dimensions.

Even within the much simplified
Hubbard or $t-J$ model close to half filling, there are two 
different energy scales in the problem. There is a piece of 
``high energy physics", which is responsible for forming local
singlet pairs. This piece of physics is not very subtle, and
basically originates from the $J$
term in the Hamiltonian. We can associate a temperature, $T_{MF}$
with this energy scale, which can be between zero and 
$1000K$, depending on the filling factor. For conventional
BCS superconductors, the energy scale of the pair formation is the
same as the true phase transition temperature into a superconducting
state. However, for the oxides, this
piece of ``high energy physics" does not tell us about the true 
ground state of the system. 

Back in 1987, Lee and Read\cite{leeread} 
asked a thought provoking question,
``Why is the $T_c$ so low?". If the singlet formation is the same
as superconducting transition, the natural transition temperature
would be the same as $T_{MF}$, much higher than the observed 
$T_c$. The reason that the $T_c$ is low is because there is 
another ``low energy physics", valid below $T_{MF}$, 
which governs the fate of the
singlet pair. The singlet pairs can form a spatially order state,
leading to a antiferromagnetic ground state, they can condense,
to form a $d$ wave superconductor, they can arrange themselves into a
spatially non-uniform state and lead to phase seperation\cite{emery2}, 
they
can form a spatially homogenous mixed state of coexisting 
antiferromagnetic and superconducting order, a ``spin bag phase"
\cite{bag2},
and finally, the effects of quantum fluctuation can be strong enough
to destroy any form of order parameter, leading to a ``resonating
valence bond (RVB) phase"\cite{anderson}.    
In order to sort out the different competing ground states, we need
a effective Hamiltonian which describes this ``low energy physics"
sector of the $t-J$ model, valid below $T_{MF}$. It should be a simple
Hamiltonian which can be solved analytically,
and yet, the complexity of the possible phases should emerge from the
unity of the model and the simplicity of its mathematical solution.

What is the principle which can guide us in the search of such a 
universal low energy Hamiltonian? The different competing orders
are not separated by any obvious energy scales in the remaining problem.
Therefore, it is not obvious how one can systematically apply the
renormalization group idea to integrate out the irrelevant degrees of
freedom. 
Fortunately, whenever Nature poses us a extraordinary 
difficult challenge involving strongly interacting low energy degrees
of freedom, it often times gives away its secret in the {\it symmetry}
of the problem.
Symmetry principle, 
exact or approximate, served us very well in our quest of
the laws of nature, and it appears to be the only principle which can
help us to solve the highly complex high $T_c$ problem. The main idea
is to identify the symmetries of the microscopic Hamiltonian, such as 
the $t-J$ model, and use the symmetry to constrain the possible form of
the low energy effective Hamiltonian.  

What symmetries should we use, and what symmetries remain to be discovered?
Well, we know the low energy effective Hamiltonian of a antiferromagnet,
it is the $SO(3)$ quantum nonlinear $\sigma$ model. Chakravaty, Halperin
and Nelson\cite{chn}
showed convincingly that the physical properties of the undoped
oxides are well described by this model. More recently, Pines and
coworkers\cite{barzykin,pines}, Chubukov, Sachdev and Ye\cite{sachdev} argued that
this model is applicable in the underdoped regime as well.
We also know the low energy 
effective Hamiltonian of a superconductor, it is described by a 
$U(1)$ quantum nonlinear $\sigma$ model, or sometimes called the $XY$
model. Doniach and Inui\cite{doniach} attempted to describe the metal insulator
transition in terms of a quantum $XY$ model, and 
recently, Emery and Kivelson\cite{emery} presented evidence that the
superconducting transition on the underdoped side of the oxides can be described by 
a renormalized classical $XY$ model.
Both the $SO(3)$ spin rotation and the $U(1)$ charge rotation
symmetries are the obvious symmetries of the microscopic $t-J$ model,
and should certainly be used as constraining symmetries of the new,
unified low energy theory.

The simplest way to construct a unified theory of antiferromagnetism 
and superconductivity  is to introduce a concept which 
I call a {\it superspin}\footnote{I would like to thank Prof. 
R.B. Laughlin for a stimulating discussion which helped me to
form this concept.}.
It is a five dimensional vector, $n_a=(n_1, n_2, n_3, n_4, n_5)$.
The first and the fifth components are the {\it super}conducting 
components of the {\it superspin}, identified with the two $d$ wave
superconducting order parameters\cite{scalapino,pines}
\begin{eqnarray}
n_1 & = & \Delta^\dagger+\Delta\ \ ,\ \  
n_5 = -i (\Delta^\dagger-\Delta)\nonumber \\
\Delta^\dagger & = & \sum_{p} g(p) c_{p\uparrow}^\dagger
c_{-p\downarrow}^\dagger\ \ ,\ \ g(p)=cosp_x-cosp_y 
\label{n15}
\end{eqnarray}
The remaining three components are the {\it spin} components of the 
{\it superspin}, identified with the antiferromagnetic order
parameter
\begin{eqnarray}
n_2 & = & \sum_{p} c_{p+Q,i}^\dagger \sigma_{ij}^x 
c_{p,j}
\ \ ,\ \ 
n_3 = \sum_{p} c_{p+Q,i}^\dagger \sigma_{ij}^y 
c_{p,j}\ \ ,\ \
n_4 = \sum_{p} c_{p+Q,i}^\dagger \sigma_{ij}^z 
c_{p,j}
\label{n234}
\end{eqnarray}
where $\sigma^\alpha$'s are the Pauli spin matrices and
$\vec Q=(\pi,\pi,\pi)$ is the antiferromagnetic ordering vector.
The concept of the superspin is similar to that of the 
pseudospin\cite{anderson2,zhang:mode},
with a crucial difference that pseudospin is really ``pseudo" in the
sense that it has no real spin component.
There is a $SO(3)$ spin symmetry acting on the $(n_2, n_3, n_4)$
subspace, with the total spin $S_{\alpha}$ being the generator of the 
rotation, and there is a $U(1)$ charge symmetry acting on the 
$(n_1, n_5)$
subspace, with the total charge $Q$ being the rotation generator,
\begin{eqnarray}
S_\alpha = \sum_{p} c_{p,i}^\dagger \sigma^\alpha_{ij} c_{p,j}\ \ ,\ \
Q = \frac{1}{2} (N-M) 
\label{SQ}
\end{eqnarray}
Here $N$ is the number of electrons and $M$ is the number of lattice
sites.
Of course, the concept of the {\it superspin} would only be useful
if we can enlarge the known $SO(3)\times U(1)$ symmetry group, to
include orthogonal transformations which can rotate the
antiferromagnetic order parameters into the superconducting ones.
The minimal group to accomplish this would be a $SO(5)$ symmetry
group. If such orthogonal transformations exists, and if they commute
with the microscopic Hamiltonian, then the concept of the 
{\it superspin} would be a very useful one. Since the orthogonal
transformations preserve the magnitude of the vector, we can impose
the constraint 
\begin{eqnarray}
n_a^2=1
\label{constraint}
\end{eqnarray}
and really think of the {\it superspin} as a vector of a fixed 
magnitude in a five dimensional space, where the antiferromagnetic
and superconducting order appear as three and two dimensional 
projections of this five dimensional vector.  
 
But we need $10$ different ``Euler angles" to describe a general 
rotation in a $5$ dimensional space, mathematically speaking, the
$SO(5)$ group has $10$ symmetry generators. (In general, the $SO(n)$
group has $n(n-1)/2$ symmetry generators). The known $SO(3)\times
U(1)$ symmetry supplies us with $4$ such generators, where are the
remaining $6$ generators?

{\bf Enter the $\pi$ operators and the $SO(5)$ algebra} 
Fortunately, the remaining $6$ symmetry generator have already been
discovered by Demler and Zhang\cite{demler}, 
in their recent work on the theory
of the resonant neutron scattering of the high $T_c$ superconductors.
These operators carry charge 2, spin 1, and a total momentum
$(\pi,\pi,\pi)$. Here we will call them $\pi$ operators, and they
are defined as follows
\begin{eqnarray}
\pi^\dagger_\alpha & = & \sum_{p} g(p) c_{p+Q,i}^\dagger 
(\sigma^\alpha \sigma^y)_{ij} c_{-p,j}^\dagger\ \ ,\ \
\pi_\alpha = (\pi_\alpha^\dagger)^\dagger
\label{pi}
\end{eqnarray}
and we see that there are obviously six of them. 
 
Let us recall that the $SO(5)$ Lie algebra has ten generators 
$L_{ab}=-L_{ba}$
satisfying the following commutation relations
\begin{eqnarray}
[L_{ab}, L_{cd}] & = & i (\delta_{ac} L_{bd} - \delta_{ad} L_{bc}
- \delta_{bc} L_{ad} + \delta_{bd} L_{ac})
\label{L}
\end{eqnarray}
Let us identify the ten operators 
$(Q, S_\alpha, \pi_\alpha, \pi_\alpha^\dagger)$ with the 
generators of the $SO(5)$, $L_{ab}$, in the following way, 
\begin{eqnarray}
L_{ab} & = & \left(
\begin{array}{ccccc}
0 &  &  &  &  \\
\pi^\dagger_x+\pi_x & 0 &  &  &  \\
\pi^\dagger_y+\pi_y & -S_z & 0 &  &  \\
\pi^\dagger_z+\pi_z & S_y & -S_x & 0 &  \\
Q & -i(\pi^\dagger_x-\pi_x) & -i(\pi^\dagger_y-\pi_y)
&-i(\pi^\dagger_z-\pi_z) & 0 
\end{array} 
\right)
\label{Lmatrix}
\end{eqnarray}
where the value of the matrix elements on the upper right triangle
are determined by antisymmetry, $L_{ab}=-L_{ba}$.
Using the fermionic representations of the 
$(Q, S_\alpha, \pi_\alpha, \pi_\alpha^\dagger)$ in equations (\ref{SQ}),
(\ref{pi}), 
and the above identification, we can check explicitly that the
$SO(5)$ commutation relations (\ref{L}) are indeed satisfied provided that
we make the continuum approximation 
\begin{eqnarray}
g^2 (p) \approx 1
\label{approx}
\end{eqnarray} 
This approximation is valid when we consider the long wave length 
limit, where the only important information about the internal
structure of a composite operator is its symmetry, but its
detailed shape. Since $g^2 (p)$ has a $s$ wave symmetry, we approximate
it by its average value.

With the aid of the $SO(5)$ algebra, the concept of the superspin
introduced earlier can be defined precisely. {\it The superspin $n_a$ 
is defined as a vector representation of the $SO(5)$ Lie
algebra}, satisfying the following commutation relation
\begin{eqnarray}
[L_{ab}, n_c] & = & -i \delta_{bc} n_a + i \delta_{ac} n_b 
\label{Ln}
\end{eqnarray} 
If we substitute for $L_{ab}$ and $n_c$
the microscopic definitions in terms of the electron operators 
(\ref{n15}), (\ref{n234}) and (\ref{Lmatrix}), 
we find that they are indeed satisfied in the continuum approximation
(\ref{approx}). In particular, we see that the $\pi$ operators rotate 
the antiferromagnetic order parameter into superconducting order
parameter and vice versa. Therefore, the six $\pi$ operators\cite{demler}
recently introduced by Demler and Zhang are exactly the six missing
symmetry generators we are searching for.

With the basic low energy variables assembled and their kinematic
relations fully specified, we are now in a position to address the
dynamic properties, and discuss the commutation relations between
the symmetry generators $L_{ab}$ and the microscopic $t-J$ 
Hamiltonian. The total spin operators $L_{23}, L_{24}, L_{34}$ 
and the total charge operator $L_{15}$ certainly commute exactly
with the $t-J$ Hamiltonian. What about the remaining six generator?

Demler and Zhang\cite{demler} have shown that although they do not
commute exactly with the Hamiltonian, they are approximate 
eigenoperators of the $t-J$ Hamiltonian, in the sense that
\begin{eqnarray}
[{\cal H},\pi^\dagger_\alpha] & = & \omega_0 \ \pi^\dagger_\alpha  
\label{eigenoperator}
\end{eqnarray} 
where $\omega_0 = J (1-n)/2 - 2\mu$.
The way to obtain this relation is by the standard $t$ matrix
approximation, similar to the random-phase-approximation 
for particle-hole operators. The reason that this relation hold
is rather simple. The $\pi^\dagger_\alpha$ operator is a particle particle 
operator, describing a pair of electrons with a center of mass
momentum $\vec Q=(\pi,\pi,\pi)$. However, for tight binding models
with nearest neighbor hopping, the two particle continuum vanishes
at $\vec Q$. Therefore, a repulsive interaction, such as $J$ in the 
triplet channel, naturally leads to a antibound state. $\omega_0$ in
the above equation is nothing but the energy eigenvalue of this antibound
state. More sophisticated calculations beyond the $t$ matrix 
approximation have also been carried out recently\cite{kohno,demler3}.
Equation (\ref{eigenoperator})
is a fundamental equation underlying the present work,
it provides a bridge from the microscopic $t-J$ Hamiltonian to
the effective nonlinear $\sigma$ model theory. One can also define
a $\pi'$ operator, using $g(p)=cosp_x+cosp_y$ in equation (\ref{pi}). 
Such a operator would
transform antiferromagnetism to $s$ wave superconductivity. However,
it was found\cite{demler} that they are in general not eigenoperators. Therfore,
our theory naturally predicts $d$ pairing symmetry.

There are two reasons why the $SO(5)$ symmetry is only a 
approximate one. One is the approximation used to derived 
equation (\ref{eigenoperator}). 
Therefore, this relation should be tested in
exact diagonalization studies of the Hubbard or the $t-J$ model.
This can be done by calculating the dynamical correlation functions
of the $\pi$ operators and identifying isolated poles in their spectral 
function.
Fortunately, such a study has been carried out recently by
Meixner and Hanke\cite{hanke}, there is indeed a sharp
collective mode in the $\pi$ channel. The second reason why the
$SO(5)$ symmetry is broken explicitly is that even if (\ref{eigenoperator})
holds exactly, the Hamiltonian does not commute with all generators
of $SO(5)$. However, this source of explicit symmetry breaking is
easy to handle, and has interesting physical consequences which 
we will explore throughout this paper. 
The eigenoperator relation (\ref{eigenoperator})
is almost as good as the vanishing
of the commutator. The reason is that the Casimir
operator $C=\sum_{a<b} L_{ab}^2$ of the $SO(5)$ algebra
commutes with the Hamiltonian, and therefore,
we can still classify all the eigenstates of the Hamiltonian 
according to their $SO(5)$ quantum numbers. This also means that the
superspin vector multiplet $n_a$ is not mixed into other higher
tensorial representations of the $SO(5)$ under time evolution,
a crucial fact which enables us to construct a low energy effective
Hamiltonian containing the superspin degrees of freedom only.
The situation is very 
analogous to the problem of a spin precessing in a magnetic field. 
While the full spin rotation symmetry is broken by the magnetic field, 
the magnitude of the spin is preserved. 

So far I have tried to present the line of arguments according to the
logical reasoning in the high $T_c$ problem. However, at this stage,
it would be appropriate to reveal the historic origin and the
source of inspiration
which lead to the present work. The present work is a generalization
of the concept of the exact $SO(4)$ symmetry\cite{so4} of the Hubbard model
and its application to the negative $U$ Hubbard model\cite{zhang:mode,zhang:so4,demler2}. 
Using the
$\eta$ operators constructed earlier by Yang\cite{yang}, Yang and 
Zhang\cite{so4}
pointed out that the Hubbard model has in additional to the usual
$SO(3)$ spin rotation symmetry also a $SO(3)$ pseudospin symmetry generated
by the following operator,
\begin{eqnarray}
\eta^\dagger = \sum_{p} c_{p+Q\uparrow}^\dagger
c_{-p\downarrow}^\dagger \ \ , \ \ 
\eta = (\eta^\dagger)^\dagger \ \ , \ \
\eta_0 = Q
\label{eta}
\end{eqnarray}
While this symmetry is valid both for the positive and the negative
$U$ Hubbard model, the simplest vector multiplet contains $s$ wave
superconducting and charge density wave order parameter
\cite{zhang:mode,zhang:so4}, 
\begin{eqnarray}
\Delta_s^\dagger = \sum_{p} c_{p\uparrow}^\dagger 
c_{-p\downarrow}^\dagger\ \ , \ \
\Delta_0 =  \sum_{p\alpha} c^{\dagger}_{p+Q,\alpha } c_{p,\alpha }\ \ , \ \
\Delta_s = (\Delta_s^\dagger)^\dagger 
\label{pseudospin}
\end{eqnarray}
and is therefore only useful for the negative $U$ model. The
simplest generalization of the pseudospin symmetry to the positive
$U$ or the $t-J$ model is the concept of the superspin introduced in 
this work, the precise correspondence between these two symmetries are
summarized in table I.  

{\bf Construction of the $SO(5)$ quantum nonlinear $\sigma$ model} 
With the basic kinematic and dynamic constraints assembled, we are in
a good position to construct a low energy effective Hamiltonian.
But before we do so, a critical reader may be concerned with the
fact that the $SO(5)$ symmetry is only approximately valid. How do
we address this question? Well, it may be useful to look at a historical
example where a approximate symmetry was used solve a important physics
problem. 
The quantum chromodynamics ($QCD$) is a model of strong interaction. 
Unlike the quantum electrodynamics, it can not be solved perturbatively.
However, in the limit of vanishing quark masses, the $QCD$ Lagrangian
has a $SU(2)\times SU(2)$ symmetry, with conserved vector and axial
currents. In real world, the quarks have small (compared to the natural
$QCD$ energy scale) but finite masses, so that the axial current is
only approximately conserved.  The idea behind the hypothesis of 
partially conserved axial current (PCAC) is to construct a effective 
nonlinear $\sigma$ model assuming that the axial currents were conserved,
and then add small explicit symmetry breaking terms to reflect the 
effects of finite quark masses. The nonperturbative, 
low energy dynamics of the $QCD$
is essentially understood in this way.
Therefore, we will follow the same strategy here to extract the low
energy content of the $t-J$ model. Along the way, we shall encounter
a pleasant surprise, although the $SO(5)$ symmetry that we start
from is approximate, near a critical point, it becomes exact under
the renormalization group flow. 

Let us first assume that the 
$SO(5)$ symmetry were a exact symmetry, and ask what is the form 
of the low energy Hamiltonian. As we argued before, below $T_{MF}$,
the system has a tendency to form local singlets, but it has not
fully decided whether to become antiferromagnetic or superconducting
in its ground state. Therefore, we can think of $T_{MF}$ as a mean
field transition below which the superspin acquires a fixed
magnitude, leaving its orientation as a low energy degrees of freedom.
Close to $T_{MF}$, the anisotropy terms in the superspin space
are not very important, and this transition can be simply described 
by a standard Landau 
free energy functional
\begin{eqnarray}
F = a|\vec n|^2 + b|\vec n|^4
\label{Landau}
\end{eqnarray}
$T_{MF}$ is the temperature below which the coefficient
of the quadratic term $a=a'(T-T_{MF})$ changes sign. 

Below $T_{MF}$, the magnitude of the superspin is fixed, we can always
rescale it to satisfy the constraint (\ref{constraint}). 
The ten generators of $SO(5)$
describe the orthogonal rotation of the superspin, leaving its 
magnitude fixed. Having either antiferromagnetism or superconductivity
corresponds to a fixed direction
of the superspin, thus breaking the $SO(5)$ symmetry spontaneously.
The low energy dynamics of such a system is determined completely
in terms of the Goldstone bosons and their nonlinear interactions
as specified by the $SO(5)$ symmetry. The kinetic energy 
of the system is simply
that of a $SO(5)$ rigid rotor, given by 
$\frac{1}{2\chi}\sum_{a<b}L_{ab}^2(x)$.
The quantity $\chi$ is nothing but the moment of inertia of the 
rigid rotor. If we assume the exact $SO(5)$ invariance, and rescale
the magnitude of the superspin to satisfy the constraint (\ref{constraint}), 
there are
no polynomial terms in the potential energy, and in the long wave length
limit,
we can perform a gradient expansion to obtain a term 
$\frac{\rho}{2} \sum_a (\partial_k n_a(x))^2$ to the leading order. Therefore,
the resulting Hamiltonian density is given by
\begin{eqnarray}
H_s = \frac{1}{2\chi}\sum_{a<b}L_{ab}^2(x) +
\frac{\rho}{2}  \sum_a (\partial_k n_a(x))^2
\label{Hs}
\end{eqnarray}

$H_s$ is a unique low energy Hamiltonian constrained by the $SO(5)$
symmetry. However, this symmetry is not exact, therefore, we should 
allow some weak symmetry breaking perturbations. Of course one should
break the $SO(5)$ symmetry in such a way that the subgroup $SO(3)\times
U(1)$ of the spin and charge symmetry is still exact. Physically, this
corresponds to going from a symmetric rotor of only one moment of inertia
to a asymmetric rotor with different moments of inertia. The asymmetric
Hamiltonian is given by
\begin{eqnarray}
H_a = \sum_{a<b} \frac{1}{2\chi_{ab}}L_{ab}(x)^2 +
\sum_{a<b} \frac{\rho_{ab}}{2} (v^k_{ab}(x))^2 + V(n)
\label{Ha}
\end{eqnarray}
where $\chi_{15}=\chi_c$, $\chi_{23}=\chi_{24}=\chi_{34}=\chi_s$ and 
$\chi_{1(2,3,4)}=\chi_{(2,3,4)5}=\chi_\pi$ are the charge, spin and the
newly introduced ``$\pi$" susceptibility, 
and we have also introduced
a generalized velocity field
$v^k_{ab}=n_a \partial_k n_b - n_b \partial_k n_a $
and the corresponding stiffness in the charge ($\rho_{15}=\rho_c$), 
spin ($\rho_{23}=\rho_{24}=\rho_{34}=\rho_s$)
and the $\pi$ ($\rho_{1(2,3,4)}=\rho_{(2,3,4)5}=\rho_\pi$)
channel. In the presence of explicit 
symmetry breaking, we also allow a quadratic symmetry breaking term of the 
form 
\begin{eqnarray}
V(n)= -\frac{g}{2} (n^2_2+n^2_3+n^2_4)
\label{V}
\end{eqnarray}

$H_s$ and $H_a$ 
describe the system at half filling only. In order
to go away from half filling, one simply has to add 
a chemical potential term $-2\mu Q= -2\mu L_{15}$, to obtain
${\cal H}_s = H_s -2\mu L_{15}$ and 
${\cal H}_a = H_a -2\mu L_{15}$ respectively. 
This Hamiltonian is to be quantized using the commutation relations
(\ref{L}) and (\ref{Ln}), and a complete set of equations of motion can be
determined.

Given the Hamiltonian ${\cal H}_a$, one can simply perform a Legendre
transformation and obtain the corresponding Lagrangian. After a simple
Wick rotation $t\rightarrow i\tau$, the Lagrangian density becomes
\begin{eqnarray}
{\cal L}_a = \sum_{a<b} \frac{\chi_{ab}}{2}\omega_{ab}(x)^2 +
\sum_{a<b} \frac{\rho_{ab}}{2} (v^k_{ab}(x))^2 + V(n)
\label{La}
\end{eqnarray}
where 
\begin{eqnarray}
\omega_{ab}=n_a (\partial_\tau n_b - i B_{bc} n_c) - (a\rightarrow b)
\label{omega}
\end{eqnarray}
is the angular velocity. Here we have introduced a set of
generalized external potentials $B_{ab}$ coupling to
$L_{ab}$. In our current problem,
the only nonvanishing component is $B_{15}=-B_{51}=2\mu$.
The partition function of the system is given by 
\begin{eqnarray}
{\cal Z} = \int [dn_a] \delta(n^2 -1) e^{-\int_0^\beta d\tau \int d^d x
{\cal L}_a}
\label{Z}
\end{eqnarray}

It is now appropriate to define the constants appearing in the model.
$\chi_s$ and $\chi_c$ are the familiar uniform spin susceptibility and
charge compressibility of the system, while $\rho_s$ and $\rho_c$
are the spin stiffness and charge stiffness respectively ($\rho_s$
should not be confused with the superfluid density which sometimes
also uses this notation). $\chi_\pi$ and $\rho_\pi$ are the two 
constants newly introduced in this work. They describe the time and
length scale over which a antiferromagnetic region can be converted to
a superconducting region and vice versa. 
In order for the $SO(5)$ symmetry to be
approximately valid, one would require these constants are close in
value. The last remaining constant of our model is the anisotropy 
constant $g$, which selects either a ``easy plane" in the superconducting
space $(n_1,n_5)$, or a ``easy sphere" in the antiferromagnetic 
space $(n_2,n_3,n_4)$, depending on the sign of $g$. As argued before,
$\mu=0$ defines the model at half filling, where we know the system
is antiferromagnetic. Therefore, we fix $g>0$ to match this fact.
At this moment, the values of the constants (especially the sign
of $g$) of the model are not deduced from any first principle 
calculations. However, once these values are fixed at half filling,
we will not allow them to vary in a arbitrary fashion. As we shall
see in the next few sections, the richness of phase diagram comes
entirely from the variation of the chemical potential. 

{\bf Origin of superconductivity}
At half filling, we choose $g>0$ so that the superspin prefers to lie
in a ``easy sphere" of $(n_2,n_3,n_4)$. 
Away from half filling, the only new term appearing in the
Hamiltonian is just the chemical potential term in 
${\cal H}_a$. 
When we investigate the Lagrangian ${\cal L}_a$, we observe 
that the chemical potential term appear as a ``gauge coupling" in the
imaginary time direction. A constant chemical potential term is a 
``pure gauge" and could therefore be ``gauged away". One's naive 
expectation would be that such a term has no dynamic consequences. 
However, one has to pay a price for doing so. Although a constant 
chemical potential term can be gauged away in the bulk, it reappears
as a twisted boundary condition in the imaginary time direction, 
and could have non-trivial consequences. 

It is more direct to investigate the Lagrangian (\ref{La}) 
with the periodic
boundary condition in the imaginary time direction. Because of the
periodic boundary condition, the classical path extremizing the path
integral are the static solutions. For the static solutions, we 
see that the only nonvanishing contribution of the kinetic energy is
the chemical potential term, which gives a effective potential
energy 
\begin{eqnarray}
V_{eff}=V(n)-\frac{(2\mu)^2}{2}(n_1^2+n_5^2)[\chi_c(n_1^2+n_5^2)
+\chi_\pi(n_2^2+n_3^2+n_4^2)]
\label{Veff}
\end{eqnarray}
Let us first take the $SO(5)$ symmetric situation where $\chi_\pi
=\chi_c=\chi$, in this case the terms in the square bracket reduces
to $\chi$.
If we were dealing with a abelian $XY$ model, such a term still does
not have any dynamical consequences. Since $n_1^2+n_5^2=1$ in such a
case, the above term reduces to a number and gives only a trivial 
shift of the ground state energy, and can not lead to any nontrivial
phase transitions. However, if the $XY$ symmetry is 
embedded into a higher symmetry group, as is done in the present case,
this term has a profound dynamic consequence.  While the $g$ term with
$g>0$ favors a antiferromagnetic ``easy sphere" $(n_2,n_3,n_4)$, 
the chemical potential term favors a superconducting ``easy plane"
$(n_1,n_5)$. This competition leads to a first order (for more detailed
discussion of the order of the transition, see a later section) 
phase transition when
\begin{eqnarray}
\mu=\mu_c=\frac{1}{2}\sqrt{g/\chi}
\label{muc}
\end{eqnarray}
For $\mu<\mu_c$, the ground state is antiferromagnetic, and the 
superconducting state has a finite energy. This energy decreases
gradually as one increases $\mu$, until a level crossing at $\mu_c$.
For $\mu>\mu_c$, the situation is reversed, the ground state is 
superconducting, and the antiferromagnetic state has a finite energy.

For $\chi_\pi \ne \chi_c$, the situation is a bit more complex. 
In the parameter regime $\chi_c>\chi_\pi$, there is always a
first order direct transition from antiferromagnetism to 
superconductivity at $\mu=\mu_c=\frac{1}{2}\sqrt{g/\chi_c}$. 
On the other hand,
in the parameter regime $\chi_c<\chi_\pi<2\chi_c$, and for 
\begin{eqnarray}
\frac{g}{\chi_\pi}<(2\mu)^2<\frac{g}{2} \frac{1}{\chi_c-\chi_\pi/2}
\label{bag}
\end{eqnarray}
there exists a intermediate ``spin bag" phase\cite{bag2} with coexisting 
antiferromagnetic and superconducting order. Schrieffer, Wen and
Zhang\cite{bag2} describe such a phase in terms of pairing the eigenstates
of the antiferromagnetic background. It is amusing to observe that
when expressed back in terms of the orginal electron operators,
their order parameter is a mixture of the antiferromagnetic,
$d$ wave superconducting order parameter {\it and} the $\pi$
operators.

The above discussion based on the Lagrangian may appear rather formal
for some readers, a intuitive physical example would be useful. Consider
a antiferromagnet with a easy axis anisotropy (say the $z$ axis). Below
the Neel transition, the Neel vector prefers to point along the easy
axis. A uniform magnetic field pointed along the easy axis creates a 
easy $xy$ plane. At a critical value of the magnetic
field, there is a ``spin flop" transition where the Neel vector changes
its orientation from the $z$ axis to the $xy$ plane. Once the underlying
$SO(5)$ symmetry is revealed, the physics of the high $T_c$ 
superconductivity is as simple as the ``spin flop" transition. We 
see that equation (\ref{omega}) describes a precession of the superspin 
about a ``fictitious magnetic field", namely the chemical potential. 
The easy axis of the antiferromagnet translates into the ``easy sphere"
$(n_2,n_3,n_4)$ while the ``easy plane" of the antiferromagnet
translates into the ``easy superconducting plane" $(n_1,n_5)$. The
transition from a antiferromagnetic ground state to a superconducting
one is that of a ``{\it superspin flop transition}". 
In fact the phase diagram of a 
easy axis antiferromagnet in the temperature and magnetic field plane
is very similar to that of a high $T_c$ superconductor in the
temperature and chemical potential plane. The complete analogy between
these two systems are summarized in the following table.

\vspace{.2in}
\begin{tabular}{|l|c|c|c|}    \hline
         &$t-J$ model  &negative $U$ Hubbard&antiferromagnet \\ 
& &model &in a $B$ field\\ \hline
symmetry &$SO(5)$      &$SO(4)$                       &$SO(3)$ \\  \hline
symmetry generators&$S_\alpha,Q,\pi_\alpha,\pi^\dagger_\alpha$&$S_\alpha,Q,\eta,\eta^\dagger$&$S_\alpha$ \\ \hline
order parameter&superspin&pseudospin&Neel vector\\ \hline
symmetry breaking&$\mu$&$\mu$&$B$\\ \hline
phase transition&superspin flop&pseudospin flop&spin flop\\ 
&(from SDW to $d$ wave &(from CDW to $s$ wave &(from easy axis to \\ 
&superconductivity)&superconductivity)&easy plane)\\ \hline
collective modes&$1$ massless phase and&$1$ massless phase and&$1$ massless $XY$ and \\ 
&$3$ massive SDW modes&$1$ massive CDW mode&$1$ massive $Z$ mode\\ \hline
\end{tabular} 
\vspace{.2in}

This analogy also helps us to understand the origin of the superspin
flop transition in the Hamiltonian formulation of the problem. In the
Hamiltonian formulation, the chemical potential term appear as a 
coupling to the symmetry generator $L_{15}$, the number operator
$Q$, but not to the superspin $n_a$ directly. It is not immediately
obvious why this term would select any particular direction in the
superspin space. Of course, the same problem
occurs in the antiferromagnetic analogy, where the uniform magnetic
field couples to the total spin vector, not the Neel vector. If one
tries to visualize the the Neel Hamiltonian in a semi-classical fashion, 
one has to remember that there is a important constraint that the
total spin vector is orthogonal to the Neel vector. The externally
applied uniform magnetic field leads to a finite total spin in the
$z$ direction. In order to satisfy the orthogonality condition, the
Neel vector therefore has to lie in the plane orthogonal to $z$. 
In the present model, (restricting to the $SO(5)$ symmetric case to
simplify the algebra), the orthogonality constraint takes the form
\begin{eqnarray}
\epsilon^{abcde} n_c L_{de}=0
\label{orthogonality}
\end{eqnarray}
which can be simply proved by expressing the angular momenta
in terms of the angular velocities 
$L_{ab}=\chi(n_a \dot n_b-n_b \dot n_a)$. The chemical
potential term leads to doping, or prefer a finite value
$L_{15}$. We immediately see that the constraint equation gives
$n_2=n_3=n_4=0$ if other $L_{ab}$ generator have no ground state
expectation value, which is the case since $\chi>0$. 

The question
of the origin of superconductivity is not one question, but rather 
two related questions separated by a energy scale. The ``high energy"
mechanism leads to binding of electrons into singlet pairs.
The origin of this binding is rather obvious in the $t-J$ model,
since the $J$ term favors electrons on near neighbor sites to 
be antiparallel. The problem is that this same $J$ could also lead
to antiferromangetism. In our new language, this pair binding gives
rise to a finite magnitude of the superspin, leaving its orientation
still unfixed. This is in marked contrast to the 
celebrated Bardeen-Cooper-Schrieffer theory of conventional
superconductivity where pair binding is equated with the onset 
of superconductivity, and also different from the large
phase fluctuation picture\cite{emery} where the only uncertainty of the
order parameter is its phase. In addition to the high energy
pair binding mechanism, a ``low energy"
mechanism selects a orientation of the superspin
and distinguishes antiferromagnetism from
superconductivity. The selection of the different possible ground states
are described by $SO(5)$ nonlinear $\sigma$ model and the mechanism for
favoring superconductivity is the ``superspin flop" mechanism 
discussed above. We see explicitly that {\it superconductivity is a
inevitable consequence of the $SO(5)$ symmetry and antiferromagnetism at
half filling}. From this point of view, a antiferromagnet state
can in some sense be thought of as a solid formed by Cooper pairs,
and the ``superspin flop" transition is a first order melting
transition from the solid into a superfluid of the Cooper pairs.
In this frame work, the ``spin fluctuation exchange"
calculation\cite{scalapino,pines}
should be interpreted as a mean field theory of $T_{MF}$.
Such calculations are important parts of our understanding, 
and lead to the glorious prediction of $d$ superconductivity,  
but they are not complete since they do not address the actual phase 
transition and the competition between antiferromagnetism and
superconductivity.  The two different mechanisms compliment each other
in their respective energy regime and together form a complete picture
of the origin of superconductivity in the oxides.

{\bf Theory of collective modes and their nonlinear interactions}
The superspin model gives a natural description of the collective
modes in the ordered phase. Close to the transition between the
antiferromagnetic and the superconducting ground states, the 
ordered phases are not conventional, but their low energy excitations
reflect the competition between the two different kinds of ordering. 
These low energy excitations can be studied systematically by 
applying symmetry principles and they have profound experimental 
consequences which we
shall discuss in this section.

Before performing concrete calculations lets us first address the 
qualitative features of the low energy excitations using symmetry
principles. Let us first suppose that the $SO(5)$ symmetry were 
exact. In this case, if the superspin has a fixed orientation in 
its ground state, this state breaks the $SO(5)$ symmetry spontaneously.   
Spontaneous breaking of a continuous symmetry naturally leads to
Goldstone bosons. The number of the Goldstone bosons is the number
of the {\it broken} symmetry generators. A fixed direction of the
superspin leaves the $SO(4)$ subgroup of the $SO(5)$ unbroken. The
$SO(5)$ group has $10$ generators, while the $SO(4)$ group has $6$
generators. Therefore, we would expect to have $4$ Goldstone bosons,
corresponding to the $4$ broken symmetry generators.

What is the nature of these 4 Goldstone bosons? Suppose that
the superspin vector
lies in the superconducting plane, say along the $n_1$ direction. The 
mode corresponding to the broken rotation generator $L_{15}$ is 
just the usual Goldstone mode, describing the phase degrees of freedom
of the superconducting order parameter. The three other modes corresponds
to the broken generators $L_{12}, L_{13}$ and $L_{14}$. These modes
form a triplet representation of the unbroken $SO(3)$ spin symmetry.
On the other hand, if the superspin vector lies in the
antiferromagnetic sphere, say along the $n_2$ direction, the usual 
spin waves corresponds to the broken generators $L_{23}$ and $L_{24}$.
They do not form a representation of the spin $SO(3)$ group, since 
this symmetry is spontaneously broken. In addition to the usual spin waves,
the current theory predicts two additional Goldstone modes, corresponding
to the broken generators $L_{21}$ and $L_{25}$. They form a 
doublet representation
of the charge $U(1)$ symmetry group. 

This symmetry analysis make it clear what happens if the $SO(5)$ 
symmetry is explicitly broken to $SO(3)\times U(1)$. 
In the presence of a quadratic symmetry breaking term 
$g_{eff}=g-\chi(2\mu)^2$, 
the spin triplet Goldstone mode of the superconducting state 
would become massive if $g_{eff}<0$
and the charge doublet mode of the antiferromagnetic state 
would become massive if $g_{eff}>0$. The usual spin wave modes and
the phase mode remain massless because of the $SO(3)\times U(1)$
symmetry. Goldstone bosons which become massive because of explicit
symmetry breaking terms are sometimes called pseudo Goldstone bosons.

This general symmetry analysis can be easily checked by a explicit
calculation starting from the ${\cal H}_a$. The equations of 
motion derived from this Hamiltonian takes the form
\begin{eqnarray}
\dot L_{ab}(x) & = & \frac{1}{2}\sum_c (\chi^{-1}_{ac}-\chi^{-1}_{bc})
                     \{L_{ac}(x),L_{bc}(x)\} +
                     \sum_c (\rho_{ac}-\rho_{bc})
                     v^k_{ac}(x)v^k_{bc}(x) \nonumber \\
               &   & + \partial_k \sum_c (\rho_{ac}n_b n_c v^k_{ac}(x)-
                                   \rho_{bc}n_a n_c v^k_{bc}(x))  +
                     \sum_c (B_{ac} L_{bc}(x) - B_{bc} L_{ac}(x))\nonumber \\
               &   & + g \sum_{\alpha=2,3,4} (n_a\delta_{b\alpha}n_\alpha-
                     n_b\delta_{a\alpha}n_\alpha) \\
     L_{ab}(x) & = & \chi_{ab} \omega_{ab}
\label{eom}
\end{eqnarray}
where $\omega_{ab}$ is given by the Wick rotation of equation (\ref{omega}). 
The first term on the
right hand side describe the {\it local} motion of a five 
dimensional asymmetrical top, a simple generalization from the equations of
motion of a asymmetrical top in classical rigid body mechanics. 
The second term arises from the asymmetrical 
coupling of the neighboring rotors. The third term is the most relevant
one for our discussion.
It arises mathematically from the Schwinger term, proportional to
the derivative of a delta function, in the local commutator
between $L_{ab}(x)$ and $v^k_{cd}(y)$, and expresses the anomaly content of
the $SO(5)$ current algebra. In the absence of the other terms on the
right hand side, there are ten conserved charges, and this term 
describes the conserved currents associated with these charges.
The fourth term simply describes the precessional motion of $L_{ab}(x)$
in the presence of a external field, and the last term describe the local
torque due to the anisotropy potential. It is easy to see that for 
$Q(x)$ and $S_\alpha(x)$, only the third term is nonvanishing, expressing the
continuity of these exactly conserved charges. 

This set of equations are the nonabelian generalization of the familiar
Josephson equations\cite{anderson3}
in conventional superconductors. The first equation is
the generalization of the 
Josephson current relation. If we take $L_{ab}(x)$ to be $L_{15}(x)$
in the first equation, we find that the only surviving term is the third
one, which reduces to the standard Josephson current expression 
$\vec J=\rho_c \vec \nabla \theta$ in the isotropic limit. The second
equation is the generalization of the Josephson acceleration equation. 
If we take $L_{ab}(x)$ to be $L_{15}(x)$ in this equation, we recover the
familiar $ac$ Josephson equation $\mu=d\theta/dt$. This set of equations
are also generalization of the Landau-Lifshitz-Neel equation of 
antiferromagnets\cite{halperin}. This statement
can be easily checked if we take $L_{ab}(x)$
to be $L_{23}(x)$, $L_{24}(x)$ and $L_{34}(x)$ in the above two 
equations.

This set of equations are the 
fundamental equations of the high $T_c$ superconductors. They 
unify the treatment of antiferromagnetim with superconductivity and 
describe the linear spectrum and the nonlinear interactions of 
the collective modes in the antiferromagnetic and the superconducting
phases.
The new physics contained in these equation can be easily analyzed 
in the linearized approximation. For $g_{eff}>0$, the ground state
is antiferromagnetic, we can linearize the equations around the
Neel vector, say $n_2$, and we obtain
\begin{eqnarray}
\chi_s \partial^2_t n_\alpha = \rho_s \partial^2_k n_\alpha
\ ,\ \alpha=3,4\ \ ; \ \
\chi_\pi \partial^2_t n_\lambda = \rho_\pi \partial^2_k n_\lambda
-(g-(2\mu)^2\chi_\pi) n_\lambda\ ,\ \lambda=1,5
\label{doublet}
\end{eqnarray}
The first equation describe the two spin wave modes of the antiferromagnet,
while the second equation predicts a new, massive doublet pairing mode. 
This mode is the precursor of superconductivity in the antiferromagnetic
phase. This mode also makes it plausible that one can think of a 
antiferromagnet as a quantum solid of Cooper pairs. A quantum solid
has two types of excitations, the gapless phonon modes, and a gapful
mode corresponding to extracting a atom from its position\cite{anderson3}. 
The pairing mode described in the above equation roughly corresponds to
the second type of excitation of a solid, if we identify the Cooper pair
as a ``atom" of the solid. We see that as the chemical potential is
increased, this pairing mode lowers its energy, the quantum solid becomes
softer, and eventually the solid melts completely to form a superfluid 
of Cooper pairs. 

For $g_{eff}<0$, the ground state is superconducting, and we can 
linearize the equations around the $n_1$ vector to obtain
\begin{eqnarray}
\chi_c \partial^2_t n_5 = \rho_c \partial^2_k n_5\ \ ; \ \
\chi_\pi \partial^2_t n_\alpha = \rho_\pi \partial^2_k n_\alpha
-((2\mu)^2\chi_\pi -g) n_\alpha\ \ , \ \
\alpha=2,3,4
\label{triplet}
\end{eqnarray}
The first equation describe the usual Goldstone mode (sound mode) of a 
superconductor,
while the second equation describe the triplet of massive magnetic
modes predicted by Demler and Zhang\cite{demler}. This mode is a precursor
of antiferromagnetism in the superconducting phase. It can be roughly 
thought of as the roton mode in a superfluid, since they both reflect
the ``diagonal short range order" in a superfluid. As the chemical 
potential is decreased, this mode lowers its energy and eventually
the superfluid ``solidifies" to form a antiferromagnet. 

Both the doublet pairing mode in the antiferromagnetic phase and
the triplet magnetic mode in the superconducting phase owe their
existance to the kinetic energy terms of the $\pi$ operators. Therefore,
we shall call them the $\pi$ doublet and the $\pi$ triplet modes
respectively. 

Both classes of new collective 
modes have important experimental consequences. The $\pi$ triplet
mode has been used by Demler and Zhang\cite{demler} 
to explain the recent resonant 
neutron scattering experiments on $YBCO$ superconductors. 
It was found that below the superconducting transition temperature,
a collective excitation peak appears in the 
neutron scattering cross section. The energies of the mode range
from $25meV$ to $41meV$\cite{neutron1,neutron2,keimer1,keimer2,dai}
depending on doping level, and the scatterings
are observed in the spin triplet channel, at the commensurate 
momentum $(\pi,\pi,\pi)$,
in exact agreement with the quantum numbers found theoretically.
These modes vanish exactly at $T_c$\cite{keimer1,keimer2}, consistent
with the interpretation that they are the pseudo-Goldstone modes  
associated with the $U(1)$ symmetry breaking.
The precise correlation between the energy of the neutron peak and
the superconducting transition temperature will be discussed in a
later section.

The $\pi$ doublet mode of a doped antiferromagnetic state is much
harder to observe experimentally. But they have indirect consequences
which can be measured. In a puzzling experiment, Suzuki 
{it et al.}\cite{geballe} observed a unusually long Josephson 
coupling length of up to $480\AA$ in the 
$YBa_2Cu_3O_7/PrBa_2Cu_3O_7/YBa_2Cu_3O_7$ junction. This experiment
could be interpreted as a signature of the small gap towards the
$\pi$ doublet pairing excitation in the insulating material
$PrBa_2Cu_3O_7$.

All above discussion assume that quantum fluctuations are not
strong enough to destroy the long range order. In the opposite limit,
when $\chi^{-1} >> \rho$, one should diagonalize the kinetic terms in 
${\cal H}_s$ and ${\cal H}_a$ first and treat the potenial energy terms
as perturbation. The kinetic terms can be diagonalized easily by 
using the representation theory of $SO(n)$. They classified by a 
integer $l$, and the Casimir operator has the value of $l(l+n-2)$
in this representation. The ground state is a $SO(5)$ singlet with $l=0$,
and the lowest energy excitation is a massive vector multiplet with $l=1$.
This multiplet is split by the anisotropy in the kinetic coefficients.
The massive pairing doublet has a energy of 
$\frac{1}{2\chi_c}+\frac{3}{2\chi_\pi}$ while the massive 
magnetic triplet has a energy of $\frac{2}{2\chi_s}+\frac{2}{2\chi_\pi}$.

{\bf The global phase diagram of high $T_c$ superconductors}
It is straightforward to determine the phase diagram of the $SO(5)$
nonlinear $\sigma$ model and compare it with the experimentally
observed phase diagram of high $T_c$ superconductors. 
The topology of the phase diagram takes different form depending on the
strength of the quantum fluctuation. In the following, we will
always be discussing the case of $d=3$, unless otherwise stated.
Our model does not have any disorder, therefore any disorder induced
phase such as spin glass is precluded in the current discussion.

Let us first discuss the so called
renormalized classical regime\cite{chn}, 
where quantum fluctuations merely
renormalize the coupling constants of the model, but they are not 
strong enough to destroy the various types of 
ordering. In this case, at half
filling, $\mu=0$ and $g>0$, the superspin lies in the ``easy sphere"
$(n_2,n_3,n_4)$ and the ground state is antiferromagnetic. 
There is a phase transition at $T_N$
to a paramagnetic phase as the temperature
is raised, and this phase transition is in the universality class
of the classical $SO(3)$ model in three dimensions. 
When the chemical potential is increased from its half
filling value of $\mu=0$, the gap towards the $\pi$ doublet 
mode decreases. Since it is easier for the superspin 
to fluctuate into other directions, 
the effective spin stiffness decreases and so does the
Neel temperature. At the critical value $\mu=\mu_c$, the superspin
has no preferential ``easy directions", and the system is described
by a isotropic $SO(5)$ nonlinear $\sigma$ model. This model has its own
critical temperature $T_{bc}$, which is in general smaller than the
Neel temperature, since the critical temperature of the $SO(N)$ model
scales inversely with $N$. If the $\mu=\mu_c$ line
is reached when $T<T_{bc}$, the system is ordered and 
the superspin defined in this limiting 
procedure still lies within antiferromagnetic ``easy sphere". 
When $\mu$ is increased beyond the critical value, the
superspin ``flops" into the superconducting plane. Since the direction
of the superspin changes abruptly at the transition, and since at the
transition point, the correlation length in the sense of Josephson
is finite, the ``superspin flop" transition 
in general first order for $T<T_{bc}$. Beyond the superspin flop
transition, the system crosses from a $SO(5)$ critical behavior over
to a $XY$ critical behavior, and one expects the superconducting transition
temperature to increase with increasing chemical potential, since the 
mass of the $\pi$ triplet mode increases. The topology of the phase
diagram in this case is depicted in Fig. (1a).

The line of the first order superspin flop transition is in general
not exactly vertical, since the susceptibilities and the stiffness in
different superspin directions may not be the same. At zero temperature
and for $\mu<\mu_c$, $Q=0$, the density of holes vanish. For
$\mu>\mu_c$, the density of holes is finite. The discontinuous jump
of the density is given by $2\chi \mu_c$. The reason for the hole
density to vanish when $\mu<\mu_c$ is because we employed a 
superspin model  with a fixed magnitude. When one relaxes the model
to a ``soft superspin" one, the hole density can be finite in this
region. However, the discontinuous jump in the density will remain
in the ``soft superspin" model. Because there is a discontinuous jump
in the density, for hole concentrations in the ``forbidden region"
$0<Q<2\chi \mu_c$, the system phase separates into a hole rich
and a hole deficient region, a possibility first pointed out by
Emery, Kivelson and Lin\cite{emery2} in the context of the $t-J$ model.
The physics in the ``forbidden density region" could be rather complicated,
and interactions not included in our current model could give rise
to various types of ordering, such as incommensurate order and ``stripe
phases". However, we believe that this types of order are not
directly related to the origin of superconductivity.

The first order superspin flop line terminates at a critical point,
from which two second order lines, one corresponding to the Neel
transition and the other corresponding to the $XY$ transition, emerge.
For this reason, this critical point is called a bi-critical point.
There are two relevant operators in the close vicinity of the
bi-critical point, the reduced temperature $t=(T-T_{bc})/T_{bc}$
and $g_{eff}$. The scaling behavior close to this point is well
studied in the literature\cite{fisher2,nelson,nelson2,fisher}. 
One of the most important result is that
even if the model defined at the short length scales is not exactly
$SO(N)$ invariant, {\it i.e.} the susceptibility and stiffness
in different directions may not be the same, the difference among them
scales to zero under the renormalization group flow when the 
bi-critical point is reached. A simple version of this result is shown
by Pelcovit and Nelson\cite{nelson}, 
who considers a $SO(N)$ model where the
the $N-th$ component has a different stiffness than the others. 
They have shown that the difference operator has negative scaling dimension
in the $2+\epsilon$ expansion of the nonlinear $\sigma$ model, and is
therefore irrelevant close to the critical point. A more general and
very elegant result was derived by Friedan\cite{friedan}, who considers a 
nonlinear $\sigma$ model in a arbitrary Riemanian manifold
with metric $g_{ab}$. Friedan
showed that the fixed point of the renormalization group flow 
is a Einstein manifold with 
$g_{ab}=\frac{1}{2\pi(d-2)}R_{ab}$, where $R_{ab}$ is the Ricci tensor. 
In particular,
if one starts with a distorted sphere $S_n$, only the average curvature
is a relevant variable, whereas the distortions preserving the
average curvature are irrelevant. 
Therefore, if one has a microscopic model without the full
spherical symmetry, a perfect sphere with constant  
curvature is produced by the large scale fluctuations close to
the critical point. This result provides a strong theoretical
justification to describe the high $T_c$ superconductors by a 
$SO(5)$ nonlinear $\sigma$ model: {\it Even if the symmetry is only
approximately valid in the microscopic models such as the
Hubbard or the $t-J$ model, close to the bi-critical point, where
the most interesting transition from a antiferromagnetic state to 
a superconducting state occurs, this symmetry becomes exact.}

Close to the bicritical point, thermodynamic quantities obey
exact scaling relations. As mentioned before, there are two
relevant operators in the problem. Finite $t$ gives a temperature
correlation length according to the scaling law
$\xi_t \sim t^{-\nu}$.
The other relevant parameter is the biquadratic symmetry breaking
term $g_{eff}$, which gives a correlation length
$\xi_g \sim g_{eff}^{-\nu_g}$
Close to the bicritical point, the singular part of the free energy
scales according
\begin{eqnarray}
f(t,g_{eff}) \sim t^{2-\alpha} \tilde f(A g_{eff}/t^\phi)
\label{scaling}
\end{eqnarray}
where $\alpha$ is the free energy exponent of the $SO(5)$ model and
$\phi$ is a crossover exponent given by
$\phi=\nu/\nu_g$.
In equation (\ref{scaling}), 
$A$ is a non-universal constant and $\tilde f$ is a universal
scaling function of a single argument.
Other singular thermodynamic quantities are be determined in a 
similar way.

The crossover exponent $\phi$ determines the way in which
the two second order lines merge into the first order line. The
universal scaling function  $\tilde f$ diverges at two arguments,
$x_+>0$ and $x_-<0$. The two second order lines are determined by
\begin{eqnarray}
g_{eff} = (x_+/A) t^\phi\ \ ,\ \ g_{eff} = (x_-/A) t^\phi 
\end{eqnarray}
Explicit calculations of the exponents can be carried out
within the large $N$ approximation\cite{hikami}. To the leading order,
$\nu=1$ and $\nu_g=1/2$, therefore $\phi=2$. From this we conclude that
the two second order lines 
merge into the first order line tangentially. (See Figure (1a)).

As we have seen before, there exist parameter regimes where 
antiferromagnetism and superconductivity can both coexist. 
In this case, the topology of the phase diagram has to be modified
to include the immediate mixed phase, the ``spin bag" phase\cite{bag2}.
The ``spin bag" phase is conceptually similar to the 
conjectured supersolid phase in $^4 H_e$.
In this case, part or the
entire first order superspin flop lines separates into two second
order lines enclosing the ``spin bag" phase. (See Figure (1b)).
The four second order
lines intersect at a tetracritical point $T_{tc}$.

So far all phase transitions discussed are classical phase transitions,
in the sense that quantum mechanical fluctuations only renormalize the
coupling constants in the problem. When 
quantum fluctuation is gradually increased, 
ordering in some phases can be
destroyed. Since the $SO(5)$ isotropic point has the largest quantum
fluctuation, one would expect $T_{bc}$ to be driven to zero first.
In this case, the bicritical point is a quantum critical point
where the antiferromagnetic state goes into the superconducting
state via a direct second order phase transition. (See Figure (1c)). 

When the quantum fluctuations are increased further, a entire region
between the antiferromagnetic phase and the superconducting phase
becomes quantum disordered. (See Figure (1d)). 
In this region, the properties of the
model are better described by first diagonalizing the kinetic energy
term of the Hamitonian. Both the triplet magnetic excitations 
and the doublet pairing excitations have finite energy gaps.
The transition from the quantum disordered phase into the 
antiferromagnetic phase is second order, 
the doublet pairing mode of the quantum disordered phase
naturally continues into the $\pi$ doublet
mode in the antiferromagnetic phase, while the triplet
magnetic excitation of the quantum disordered phase
becomes the two gapless Goldstone modes 
of the antiferromagnetic phase. 
The transition from the quantum disordered phase into the 
superconducting phase is second order as well,
the triplet magnetic excitation of the quantum disordered phase 
naturally continues into the $\pi$ triplet
mode of the superconducting phase,\footnote{I would like to thank
Prof. T.M. Rice for a stimulating discussion which lead me to this
identification.}
while the doublet pairing mode becomes the gapless Goldstone mode
of the superconductor.

The $SO(5)$ quantum disordered phase differ from the $SO(3)$
quantum disordered phase of the quantum antiferromagnet\cite{chn} and the
quantum $XY$ disordered phase\cite{doniach} 
since it has both a massive triplet 
magnetic excitation and a massive doublet of pairing excitation.
It is tempting to identify it with 
the resonating-valence-bond (RVB) phase
introduced by Anderson. In fact, the type (1d) phase diagram
looks very similar to the original RVB idea proposed by
Anderson\cite{anderson}. 
The basic idea is that with low dimensionality and
increased doping, quantum fluctuations will first ``melt" the
antiferromagnetic state to form a resonating valence bond
liquid of spin singlet pairs, and these singlet pairs further
condense into a superfluid state. 
However, the $SO(5)$ quantum disordered
phase differ from Anderson's $RVB$ phase in terms of the excitation
spectrum. In the $RVB$ phase, spinons and holons are the
elementary excitations, while in the $SO(5)$ quantum disordered
phase, even the electrons are ``bound" into the collective coordinates.
In some sense, the $SO(5)$ quantum disordered phase describe a 
``incompressible liquid" of singlet pairs, 
similar to the Laughlin liquid in the
fractional quantum Hall effect.

We therefore see that the global features of the
phase diagram deduced from the
$SO(5)$ quantum nonlinear $\sigma$ model agree reasonablly well with
the general topology of the 
experimentally observed phase diagram of high $T_c$
superconductors. With this model, we can understand 
the depression of the Neel and the superconducting temperature 
in terms of a common origin, namely a region in the phase diagram
where the $SO(5)$ fluctuations are maximal. This theory makes precise
predictions about the scaling behavior in the crossover region. These
predictions can be tested experimentally in sufficiently clean
systems where a direct first order transition becomes possible. 
I believe that phase diagrams of the type (1a) or (1c) are 
reasonablly close to the reality in the high $T_c$ superconductors.
Type (1b) may be useful to understand the heavy fermion materials,
while type (1d) could be realized in the quasi one dimensional
ladder systems where quantum fluctuations are strong.

{\bf Theory of the spin gap, the superconducting transition
and their relationship}
One of the most interesting phenomenon of the underdoped high
$T_c$ materials is the so called spin gap. Above a certain temperature
$T_s$, the magnetic correlation length increases with decreasing
temperature, while it saturates below $T_s>T_c$. This phenomenon has
only been observed in the $YBCO$ family of the high $T_c$ superconductors.
The spin gap phenomenon
has been observed in two different kinds of nuclear spin relaxation
processes, the $1/T_1$ and $1/T_{2G}$ measurements respectively.
Through a series of insightful phenomenological analysis of the 
$NMR$ experiments, Sokol and Pines\cite{sokol} concluded that this
two experiments can be linked through a simple $z=1$ dynamical 
scaling hypothesis. Since our model is Lorentz invariant, their
hypothesis is valid here. Therefore, we shall only address here the
$1/T_{2G}$ experiment, which meausures the static spin correlation
length.

The phenomenon of spin gap has a natural explanation within the current
theoretical model. There are three different temperature scales in
the current model. When the temperature is lowered below $T_{MF}$, 
the superspin vector acquires a finite magnitude. 
However, for a finite range of temperature below $T_{MF}$, 
say $T_s<T<T_{MF}$, (see Fig. 1), the temperature
scales is still high enough so that the superspin does not ``notice" the
anisotropy is its orientational degrees of freedom. In this range of
temperature, the model is essentially $SO(5)$ symmetric, and the 
antiferromagnetic correlation length increases together with the
pairing correlation length as temperature is lowered. 
For $T<T_s$, the thermal energy becomes  
low compared to the anisotropy energy in the superspin space, 
and the superspin vector lies preferably in the superconducting plane.
Therefore, below $T_s$, the antiferromagnetic correlation length 
saturates to a finite value. Finally, at $T_c$, the superspin vector
picks a particular direction within the superconducting plane, therefore
breaking the $U(1)$ gauge symmetry, and the system becomes 
superconducting. A very similar picture applies to the antiferromagnetic
side of the transition. There is a ``pairing gap" temperature $T_p$,
above which the
pairing correlation length increases together with the antiferromagnetic
correlation length, while it saturates below $T_p$.
    
Both $T_s$ and $T_c$ can be calculated quantitatively within the 
$SO(5)$ nonlinear $\sigma$ model. In principle, one can perform the
calculation for all four types of phase diagrams shown in Fig. 1. Due
to the space limitation, we will only show the calculation for a
representative case, the case (1c), where the transition from 
antiferromagnetism to superconductivity occurs via a zero temperature
quantum critical point. Since the anisotropy is irrelevant near the
critical point, we will take the model to be isotropic. The idea 
of the large $N$ mean field theory is to implement the constraint
in the partition function (\ref{Z}) 
through a Lagrangian multiplier $\lambda$
and determine the value of $\lambda$ through a self-consistency
condition that $n_a^2=1$ is enforced on average. Skipping the intermediate
steps\cite{sachdev}, the resulting self-consistency condition is given
by
\begin{eqnarray}
2 \chi & = &
2 \int_0^{a^{-1}}\frac{d^3q}{(2\pi)^3}\frac{1}{\sqrt{c^2 q^2 + c^2 \xi^{-2}}\ 
tanh(\frac{\beta}{2}\sqrt{c^2 q^2 + c^2 \xi^{-2}})} \nonumber \\
& & + 3 \int_0^{a^{-1}} \frac{d^3q}{(2\pi)^3}
\frac{1}{\sqrt{c^2 q^2 + c^2 \xi^{-2}+\omega_0^2}\ 
tanh(\frac{\beta}{2}\sqrt{c^2 q^2 + c^2 \xi^{-2}+\omega_0^2})} 
\label{consistency}
\end{eqnarray}
Here $a$ is a lattice cutoff, $\xi$ is the superconducting correlation
length, $\omega_0^2=(2\mu)^2-(2\mu_c)^2$ 
is the energy of the $\pi$ triplet mode
in the superconducting state, $\xi_s^{-2} = \xi^{-2} + (\omega_0/c)^2$
is the antiferromagnetic correlation length and $c=\sqrt{\rho/\chi}$
is the sound or spin wave velocity.

The quantum bicritical point is determined by the condition that
at $\omega_0=0$, $\xi$ diverges at zero temperature. This determines
\begin{eqnarray}
\sqrt{\rho\chi} = \frac{5}{8\pi^2} a^{-2}
\label{qcp}
\end{eqnarray}
to be the condition for the quantum bi-critical point. 
When $\omega_0$ is greater than zero, $\xi$ diverges at a finite
temperature $T_c$. The relationship between these two quantities can
be easily determined from equation (\ref{consistency}), 
and is found to be
\begin{eqnarray}
\frac{\omega_0^2}{(kT_c)^2} (\frac{3}{2}x^2-3f(x)) \approx 16.44
\label{Tc}
\end{eqnarray}
where $x=ca^{-1}/\omega_0=\xi_s/a$ 
is the antiferromagnetic length at zero temperature,
and 
$f(x)=\frac{1}{2}x\sqrt{x^2+1}-\frac{1}{2}ln(x+\sqrt{x^2+1})$.
We see here that up to a logarithmic term, 
$\omega_0$ and $kT_c$ are simply proportional to each other.
This fact can be understood without any explicit calculations. 
At $\omega_0=0$, the correlation length at a finite temperature is
simply given by the Lorentz transformation of the finite width $\beta$
in the imaginary time direction, or the thermal de Broglie wave length
$\hbar c/kT$. At $T=0$ and finite $\omega_0$, the correlation length
can be estimated from power counting the scaling dimension of $\omega_0$,
which gives $c/\omega_0$. Equating these two length scales and 
neglecting the logarithmic corrections gives the estimate of 
$kT_c \propto \omega_0$. 

Let us now turn to the calculation of the spin gap temperature $T_s$. 
From the form of the antiferromagnetic correlation length
$\xi_s^{-2} = \xi^{-2} + (\omega_0/c)^2$, 
(a very similar form was first used by Millis, Monien and 
Pines\cite{mmp} in their phenomenological analysis of the NMR data),
we see that at high temperature,
the second term is smaller than the first, therefore, $\xi_s$
depends on temperature in the same way $\xi$ does, namely proportional
to the thermal de Broglie wave length $\hbar c/kT$. However, at low
temperature, the second term dominates, and $\xi_s$ saturates to 
a finite, temperature independent value. The spin gap temperature
can therefore be defined by $c^2 \xi^{-2} (T_s) = \omega_0^2$.
Inserting this definition into equation (\ref{consistency}), 
we obtain a explicit
formula for the spin gap temperature
\begin{eqnarray}
\frac{\omega_0^2}{(kT_s)^2} 
(\frac{5}{2}x^2-2f(x)-6f(x/\sqrt{2})) \approx 16.44
\label{Tsg}
\end{eqnarray}
Equations (\ref{Tc}) and (\ref{Tsg}) relates three experimentally observable
quantities, the superconducting transition temperature $T_c$, 
the spin gap temperature $T_s$ and the resonant neutron scattering
peak $\omega_0$ to each other. $\xi_s$ is the zero temperature
antiferromagnetic correlation length averaged over three space dimensions.
Since there are some uncertainies about the correlation length in the
$c$ axis and the value of the microscopic cutoff of the $\sigma$ model,
we shall take $\xi_s/a$ as a fitting parameter.

The relation between the energy of the resonant neutron scattering
peak and $T_c$ has been measured 
recently in the $YBa_2Cu_3O_{6+x}$ superconductors
by Fong {\it et al}\cite{keimer2}. It was found that $\omega_0$ is
linearly proportional to $T_c$ with a slope of about $5.6$. If we
take $\xi_s/a$ to be $0.8$ in equation (\ref{Tc}) and roughly doping
independent in the doping regimes explored in experiment, 
our theory gives the same slope. Notice that since $\xi_s/a$
is a three dimensional average, the actual antiferromagnetic 
correlation length in the two dimensional sheet could be much greater.

Combining the result of this section and the previous section on
the collective modes, we see that the current work offers a unique 
explanation of the recent neutron scattering
experiment of the high $T_c$ superconductors in terms of the 
pseudo-Goldstone bosons associates with the spontaneous $SO(5)$ symmetry 
breaking. Since these magnetic excitations occur only below the 
superconducting
transition temperature, the interpretation in terms of the Goldstone
bosons is most natural, and we see that the linear proportionality
between $T_c$ and $\omega_0$ is also naturally explained within the
current theoretical model. A class of theoretical 
models\cite{keimer1,bulut,mazin,liu,barzykin,millis} attempt to associate
the neutron resonance peak with the opening up of a $d$ wave pairing
gap, and possible excitonic states inside this gap. At the time when
this class of theories were proposed, only a $41meV$ peak was observed
in the optimally doped $YBCO$ 
superconductor\cite{neutron1,neutron2,keimer1}.
However, recent photoemission\cite{shen}
and resonant neutron scattering\cite{keimer2,dai} experiments in
the underdoped regime can not be reconciled with this class of theories.
In the underdoped regime, the electronic energy gap opens up at a 
temperature much higher than $T_c$. If the resonant neutron scattering
peak arises simply from the removal of spectral weight due to the 
electronic gap opening, such a peak would be observed far above $T_c$,
in contradiction to the experimental observation\cite{keimer2,dai}. In this
class of theories, the energy of the neutron scattering peak would also be
proportional to the electronic gap energy. As the doping level 
is decreased
from the optimal value, the electronic energy gap actually increases
slightly\cite{shen2}. Therefore, this class of theories would predict
that the energy of the neutron resonance increases with decreasing
doping, also in 
contradiction with the recent experiment by 
Fong {\it et al.}\cite{keimer2}.

This value of  $\xi_s/a \approx 0.8$ can now be used to estimate 
the spin gap temperature without any additional fitting parameter.
From equation (\ref{Tc}) and (\ref{Tsg}), 
we find $T_s/T_c\approx 1.37$.
There are some experimental uncertainties about the value of $T_s$
as well. The most acurate measurement of $T_s$ is in 
the $T_c=62K$ material $YBa_2Cu_3O_{6.63}$. The 
$1/T_{2G}$ measurement shows that the antiferromagnetic correlation
length saturate at $T_s\approx 100K$\cite{takigawa}. This gives a 
$T_s/T_c$ ratio of $1.61$, about $15\%$ higher than the 
theoretically predicted ratio. 

We would like to remark that these quantitative calculations are 
strictly speaking only valid near the critical doping, while most
of the experiments quoted above are far from it. It would be highly
desirable to have sufficiently clean systems close to the critical
doping where precise
comparison with theory can be made.

{\bf Attempt at a synthesis} The most important message from this work
is that antiferromagnetism and superconductivity are the two
different sides of the same coin. They are intimately related by a 
$SO(5)$ symmetry which determines the competition and the low energy
dynamics of the high $T_c$ superconductors. The current theory offers
a unified explanation of some of the most challenging problems of
high $T_c$ superconductivity.

This work answers a number of experimental puzzles. The mechanism 
of superconductivity is devided into a ``high energy" piece, 
identified with the pair binding, or the formation of the amplitude
of the superspin and a ``low energy" piece, involving a ``superspin
flop" mechanism which selects the orientation of the superspin, 
and resolves the competition between superconductivity and 
antiferromagnetism. These two different energy scales have different
doping dependence. With increasing doping, $T_{MF}$ decreases
because the effective $J$ decreases, but $T_c$ increases because
both the thermal and the quantum $SO(5)$ fluctuations decreases.
Recent photoemission experiment\cite{shen2}
indeed found that the electronic gap decreases with increasing
$T_c$, a fact inconsistent with the weak coupling BCS theory, 
but consistent with the current model. 
This work also answers the question of why $T_c$ is low compared with
the energy scales of the pair formation. Emery and Kivelson\cite{emery} 
rightfully argues that
this happens because the superfluid density is low, but did not offer a
explanation of why this is so. The current work explains this fact in
terms of increased quantum and classical $SO(5)$ fluctuations near
the isotropic point $\mu_c$, so that both the Neel temperature and the
superconducting transition temperature are depressed in this transition
region. The puzzeling spin gap phenomenon is naturally interpreted
in terms of the competition between entropy and anisotropy energy
in the superspin space. So far the most direct experimental
evidence for the approximate $SO(5)$ symmetry is the resonant neutron
scattering peaks in the underdoped and optimally doped region. 
These modes are naturally interpreted in terms of the pseudo Goldstone
bosons associated with the spontanuous $SO(5)$ symmetry breaking. Current
model predicts that their intensity vanish exactly at $T_c$, the 
mode energy is temperature independent, and decreases with decreasing
doping. All these facts are consistent with the experimental findings.
Notablly absent in our discussion is the transport properties of the
high $T_c$ oxides. Below $T_{MF}$, transport properties could be 
addressed by considering fermions coupled to the $SO(5)$ nonlinear
$\sigma$ model degrees of freedom. However, quantitative details have
not yet been fully worked out. 

The current theory also unifies a number of seeeming divergent
theoretical approaches to the high $T_c$ problem. It outlines a
strategy to systematically extract the low energy content of the
$t-J$ model by constructing low energy field theory constrained by
the microscopic symmetry. The $SO(5)$ symmetry can be used as a
basic principle to organize the various theoretical proposals.
Most directly, it unifies the $SO(3)$ nonlinear $\sigma$ model
theory\cite{chn,pines,sachdev} with the $U(1)$ nonlinear $\sigma$ model
theory\cite{doniach,emery}. 
The ``spin fluctuation exchange"\cite{scalapino,pines}
approach should be interpeted as a theory of $T_{MF}$. the
phase separation in the $t-J$ model\cite{emery2}
occurs because the ``superspin flop"
transition is first order, therefore giving rise to a ``forbidden
density region". The ``spin bag phase"\cite{bag2} can emerge in the
phase diagram as a result of increased $\pi$ fluctuation, or
more precisely when $\chi_\pi>\chi_c$. The $SO(5)$ quantum disordered
phase could in some sense 
be associated with the $RVB$ phase in Anderson's\cite{anderson}
original proposal, and occurs as a result of increased $SO(5)$
quantum fluctuation.   

The $SO(5)$ theory makes a number of new experimental predictions.
The most direct prediction is that when the materials are sufficiently
clean, there could be a direct first order transition between the 
antiferromagnetic and the superconducting states. Measurements close
to the bicritical point can be compared quantitatively with the
theoretical predictions and can serve as important test of the 
theory. There are other predictions,
due to the space limitations I shall only outline the basic ideas,
and publish the details elsewhere. In the conventional Landau-Ginzburg
theory of superconductivity, the order parameter is constrained to
lie in a plane. Near the core of a superconducting vortex, a mathematical
singularity is unavoidable. However, in the superspin model, since the
order parameter is five dimensional, the superspin could lie in
the superconducting plane far away from a vortex, but flip into the
antiferromagnetic sphere inside a vortex. Such a topological
configuration is called a ``meron" in the field theory literature, 
meaning half of a Skyrmion. Therefore, the current theory would predict
that the core of a vortex in underdoped superconductors are not
filled with normal electrons but are antiferromagnetic instead. 
Such a effect could
be observed in the dissipation dynamics, or more directly, by studying
the form factors of the vortex lattice in neutron scattering experiments.
The current theory also predicts that at the interface between 
a antiferromagnet and a superconductor, the massive $\pi$ modes
become gapless, in very much the same way the collective modes in
a Laughlin liquid in the quantum Hall effect become gapless at the 
edge. Such effects could be observed in tunneling experiments.  

High $T_c$ superconductivity is one of the most complex and perplexing
phenomena ever encountered in condensed matter physics. Obviously, if
there is any hope in completely solving this problem, the solution has
to be derived from a simple and unifying principle. The present work is
a attempt towards a synthesis of the complex high $T_c$
phenomenology from a symmetry
principle which unifies antiferromagnetism with superconductivity. 
Throughout the history of physics, symmetry principle has been a 
faithful companion in our quest of the deep unity and harmony of Nature.
May this time honored, deeply aesthetic principle help us one more time,
to unlock the secret of high $T_c$ superconductivity.

{\bf Acknowledgement} I would like to thank Prof. J. R. Schrieffer, 
S. Kivelson, R.B. Laughlin and D. Scalapino for patiently teaching me about
various aspects of the high $T_c$ problem throughout the years,  
and for sharing with me their
deep insights and enthusiasm. I would like to thank Prof. C.P. Burgess and 
C. L\"{u}tken for tutoring me about the nonlinear $\sigma$ model theory of
$QCD$. Inspiring discussions with N. Bulut, E. Demler, H. Fukuyama, 
W. Hanke, B. Keimer, H. Kohno, S. Meixner, N. Nagaosa and Z.X. Shen are also
gratefully acknowledged. Last but not least, I would like to thank
Prof. C.N. Yang for collaborating on a earlier work which inspired the
current one, and for impressing me with the power of symmetry.

This work is supported by the NSF under grant numbers DMR-9400372 and
DMR-9522915. Part of this work is carried out while visiting the IBM
Almaden Research Center and I would like to thank Dr. B.A. Jones for
the hospitality extended to me during my visit.

%\bibliography{hightc}

\begin{thebibliography}{10}

\bibitem{anderson}
P.W. Anderson.
\newblock {\em Science}, 235:1196, 1987.

\bibitem{zhangrice}
F.C. Zhang and T.M. Rice.
\newblock {\em Phys. Rev. B}, 37:3759, 1988.

\bibitem{leeread}
P.A. Lee and N.~Read.
\newblock {\em Phys. Rev. Lett.}, 58:2891, 1987.

\bibitem{emery2}
V.J. Emery, S.~Kivelson, and H.Q. Lin.
\newblock {\em Phys. Rev. Lett.}, 64:475, 1990.

\bibitem{bag2}
J.R. Schrieffer, X.G. Wen, and S.C. Zhang.
\newblock {\em Phys. Rev. B}, 39:11663, 1989.

\bibitem{chn}
S.~Chakravarty, B.I. Halperin, and D.R. Nelson.
\newblock {\em Phys. Rev. B}, 39:2344, 1989.

\bibitem{barzykin}
V.~Barzykin and D.~Pines.
\newblock {\em Phys. Rev. B}, 52:13585, 1995.

\bibitem{pines}
D.~Pines.
\newblock {\em Physica C}, 235:113, 1994.

\bibitem{sachdev}
A.V. Chubukov, S.~Sachdev, and J.~Ye.
\newblock {\em Phys. Rev. B}, 49:11919, 1994.

\bibitem{doniach}
S.~Doniach and M.~Inui.
\newblock {\em Phys. Rev. B}, 41:6668, 1990.

\bibitem{emery}
V.J. Emery and S.~Kivelson.
\newblock {\em Nature}, 374:434, 1995.

\bibitem{scalapino}
D.~Scalapino.
\newblock {\em Phys. Rep.}, 250:329, 1995.

\bibitem{anderson2}
P.W. Anderson.
\newblock {\em Phys. Rev.}, 112:1900, 1958.

\bibitem{zhang:mode}
S.~C. Zhang.
\newblock {\em Phys. Rev. Lett.}, 65:120, 1990.

\bibitem{demler}
E.~Demler and S.C. Zhang.
\newblock {\em Phys. Rev. Lett.}, 75:4126, 1995.

\bibitem{kohno}
H.~Kohno, B.~Normad, and H.~Fukuyama.
\newblock {\em Proceedings of 10th Anniversary HTS Workshop}.
\newblock World Scientific Publishing Company.

\bibitem{demler3}
E.~Demler and S.C. Zhang.
\newblock to be published.

\bibitem{hanke}
S.~Meixner and W.~Hanke.
\newblock to be published.

\bibitem{so4}
C.N. Yang and S.C. Zhang.
\newblock {\em Mod. Phys. Lett. B}, 4:759, 1990.

\bibitem{zhang:so4}
S.~C. Zhang.
\newblock {\em Int. J. Mod. Phys. B}, 5:153, 1991.

\bibitem{demler2}
E.~Demler, S.C. Zhang, N.~Bulut, and D.~Scalapino.
\newblock {\em Int. J. Mod. Phys. B}, 10:2137, 1996.

\bibitem{yang}
C.N. Yang.
\newblock {\em Phys. Rev. Lett.}, 63:2144, 1989.

\bibitem{anderson3}
P.W. Anderson.
\newblock {\em Basic Notions of Condensed Matter Physics}.
\newblock The Benjamin/Cummings Publishing Company.

\bibitem{halperin}
B.I. Halperin and P.C. Hohenberg.
\newblock {\em Phys. Rev.}, 188:898, 1969.

\bibitem{neutron1}
J.~Rossat-Mignod et~al.
\newblock {\em Physica C}, 185:86, 1991.

\bibitem{neutron2}
H.~Mook et~al.
\newblock {\em Phys. Rev. Lett.}, 70:3490, 1994.

\bibitem{keimer1}
H.F.~Fong et~al.
\newblock {\em Phys. Rev. Lett.}, 75:316, 1995.

\bibitem{keimer2}
H.F.~Fong et~al.
\newblock Superconductivity induced anomalies in the spin exitation spectra of
  underdoped $YBa_2Cu_3O_{6+x}$.
\newblock preprint.

\bibitem{dai}
P.~Dai et~al.
\newblock Magnetic dynamics in underdoped $YBa_2Cu_3O_{7-x}$.
\newblock preprint.

\bibitem{geballe}
Y.~Suzuki et~al.
\newblock {\em Phys. Rev. Lett.}, 328:328, 1994.

\bibitem{fisher2}
M.E. Fisher and D.R. Nelson.
\newblock {\em Phys. Rev. Lett.}, 32:1350, 1974.

\bibitem{nelson}
R.A. Pelcovits and D.R. Nelson.
\newblock {\em Phys. Lett. A}, 57:23, 1976.

\bibitem{nelson2}
D.R. Nelson and R.A. Pelcovits.
\newblock {\em Phys. Rev. B}, 16:2191, 1977.

\bibitem{fisher}
D.S. Fisher.
\newblock {\em Phys. Rev. B}, 39:11783, 1989.

\bibitem{friedan}
D.H. Friedan.
\newblock {\em Annals of Physics}, 163:318, 1985.

\bibitem{hikami}
S.~Hikami and R.~Abe.
\newblock {\em Prog. Theor. Phys.}, 52:369, 1974.

\bibitem{sokol}
A.~Sokol and D.~Pines.
\newblock {\em Phys. Rev. Lett.}, 71:2813, 1993.

\bibitem{mmp}
A.J. Millis, H.~Monien, and D.~Pines.
\newblock {\em Phys. Rev. B}, 42:167, 1990.

\bibitem{bulut}
N.~Bulut and D.~Scalapino.
\newblock {\em Phys. Rev. B}, 53:5149, 1996.

\bibitem{mazin}
I.I. Mazin and V.M. Yakovenko.
\newblock {\em Phys. Rev. Lett.}, 75:4134, 1995.

\bibitem{liu}
D.Z.~Liu et~al.
\newblock {\em Phys. Rev. Lett.}, 75:4130, 1995.

\bibitem{millis}
A.~Millis and H.~Monien.
\newblock On the bilayer coupling in the yttrium-barium family of high
  temperature superconductors.
\newblock preprint.

\bibitem{shen}
A.G.~Loeser et~al.
\newblock {\em Science}, 273:325, 1996.

\bibitem{shen2}
J.M.~Harris et~al.
\newblock Anomalous superconducting state gap size vs. $t_c$ behavior in
  underdoped $Bi_2Sr_2Ca_{1-x}Dy_xCu_2O_{8+\delta}$.
\newblock preprint.

\bibitem{takigawa}
M.~Takigawa.
\newblock {\em Phys. Rev. B}, 49:4158, 1994.

\end{thebibliography}
%\bibliographystyle{unsrt}

\begin{figure}
\caption{Global phase diagram of high $T_c$ superconductors in the
absence of disorder. $T_{MF}$ is
the temperature below which electrons bind to form singlet pairs, in the
current model, it corresponds a finite magnitude of the superspin. 
$T_N$ and $T_c$ are the Neel and superconducting transition temperatures
respectively. 
There are four possible types of transitions from the antiferromagnetic 
to the superconducting state. a) There is a direct first order transition
which terminates at a bi-critical point $T_{bc}$. The first order
transition can be described as a ``superspin flop" transition. 
b) There are two second order phase transitions with a 
intermediate ``spin bag"
phase. The four second order lines merge at a tetracritical point
$T_{tc}$. c) There is a single second order phase transition at
a quantum critical point. d) There are two second order quantum phase
transitions with a intermediate quantum disordered phase.
In each type of the phase diagrams, there are two crossover temperatures
$T_s$ and $T_p$. The ``spin gap" temperature $T_s$ corresponds to
the temperature below which the superspin lies within the
the superconducting plane, and ``pairing gap" temperature $T_p$ 
corresponds to the temperature below which the superspin lies 
within antiferromagnetic sphere.}
\end{figure}
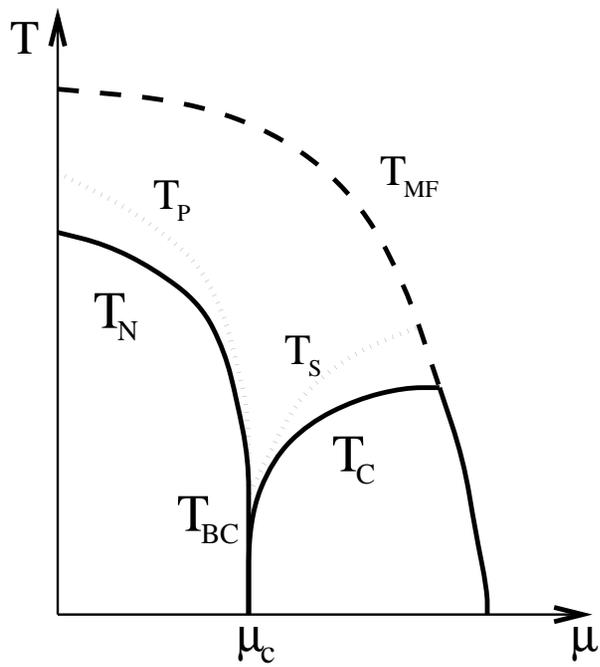
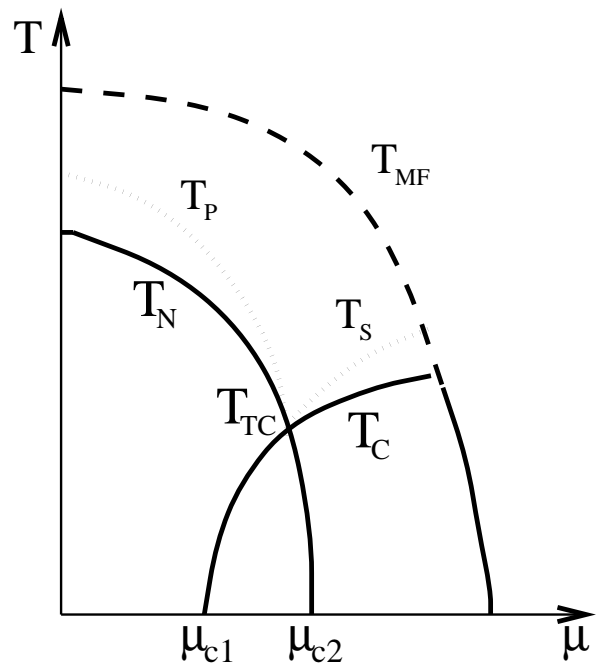
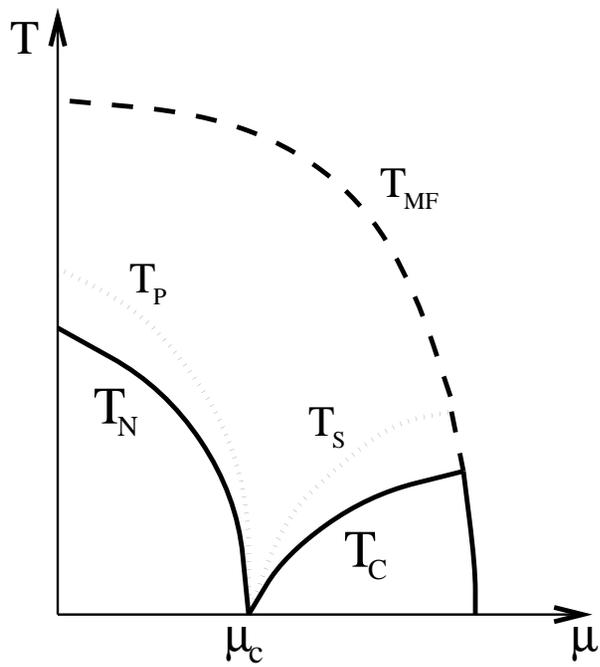
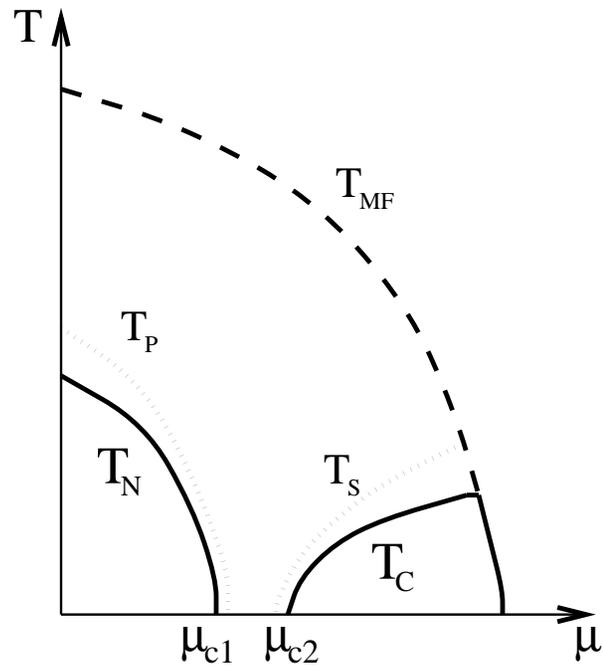
\end{document}